\documentclass[12pt]{article}

\def\mass{\mathcal M}

\def\cala{\cal A}
\def\A{{\cal A}}
\def\gzero{\bar g_{ab}}

\def\szero{\bar s_{ab} }
\def\pizero{\bar\pi^{ab} }
\def\M{{\cal M}}

\def\L{{\cal L}}
\def\P{{\cal P}}
\def\calp{\mathcal{P}}
\def\calm{\mathcal{M}}
\def\calt{\mathcal{T}}
\def\call{\mathcal{L}}
\def\cala{\mathcal{A}}

\def\ab{ab}

\begin{document}

\begin{titlepage}
\vfill
\begin{flushright}
\end{flushright}

\vfill
\begin{center}
\baselineskip=16pt
{\Large\bf The First Law for Boosted Kaluza-Klein}
\vskip 0.5cm
{\Large\bf  Black Holes}
\vskip 1.0cm
{\large {\sl }}
\vskip 10.mm
{\bf David Kastor, Sourya Ray and Jennie Traschen} \\
\vskip 1cm
{

       Department of Physics\\
       University of Massachusetts\\
       Amherst, MA 01003\\
}
\vspace{6pt}
\end{center}
\vskip 0.5in
\par
\begin{center}
{\bf Abstract}
 \end{center}
\begin{quote}
We study the thermodynamics of Kaluza-Klein black holes with momentum along the compact dimension, but vanishing angular momentum.  These black holes are stationary, but non-rotating.
We derive the first law for these spacetimes and find that the  parameter conjugate to variations in  the length of the compact direction is an effective tension, which generally differs from the ADM tension.  For the boosted black string, this effective tension is always positive, while the ADM tension is negative for large boost parameter.    We also derive two Smarr formulas, one that follows from time translation invariance, and a second one that holds only in the case of exact translation symmetry  in the compact dimension.  Finally, we show that the `tension first law' derived by Traschen and Fox in the static case has the form of a thermodynamic Gibbs-Duhem relation and give its extension in the stationary, non-rotating case.
\vfill
\vskip 2.mm
\end{quote}
\end{titlepage}

\section{Introduction}

The physics of Kaluza-Klein black holes, {\it i.e.} black hole spacetimes asymptotic at infinity to $M\times S^1$, has proved to be a surprisingly rich subject, including such phenomena as the Gregory-Laflamme instability, non-uniform static black strings and the black hole/black string phase transition  (see {\it e.g.} the reviews \cite{Kol:2004ww,Harmark:2007md}).  Research to date has focused primarilly on the static case.  However, it is also of interest to explore the properties of stationary solutions.   
Accordingly, in this paper we will study the thermodynamics of  stationary Kaluza-Klein black holes\footnote{Aspects of stationary Kaluza-Klein black holes have been studied in \cite{Hovdebo:2006jy}\cite{Kleihaus:2007dg}.  The thermodynamics of asymptotically AdS, boosted domain walls have been investigated in reference  \cite{Cai:2003xv}.}.  

Static Kaluza-Klein black holes are characterized at infinity by the mass $\calm$, tension $\calt$ and the length $\call$ of the compact direction.  The physical meaning of the tension follows from its role in the first law for static $S^1$ Kaluza-Klein black holes \cite{Townsend:2001rg}\cite{Harmark:2003eg}\cite{Kastor:2006ti}
\begin{equation}\label{static-first-law}
d\calm = {\kappa\over 8\pi G}d\cala +\calt d\call.
\end{equation}
We see that the tension determines the variation of the mass with varying length of the compact direction, under the constraint that the horizon area is held fixed.  Within the thermodynamic analogy, it appears to be an intensive parameter of the system, like temperature or pressure.

Stationary Kaluza-Klein black holes can carry linear momentum in the compact direction, as well as angular momentum.      In this paper, we will be interested in this linear momentum, which we denote by $\calp$, and will assume that the angular momentum vanishes.  The simplest solutions with 
$\calp\neq 0$ are boosted black strings.  These are obtained by starting from the infinite uniform black string, boosting in the $z$ direction and then identifying the new $z$ coordinate with period $\call$.  The boosted black string is then locally, but not globally, the same as  the static uniform black string.  
Further stationary, but not $z$-translationally invariant, solutions may be obtained by giving localized black holes or non-uniform black strings velocity in the compact direction.  

In subsequent sections, we present the following results.  We use the Hamiltonian methods of \cite{Sudarsky:1992ty}\cite{Traschen:1984bp}\cite{Kastor:2006ti} to establish the first law for stationary, non-rotating Kaluza-Klein black holes.  We also derive two Smarr formulas for these spacetimes.  These
are exact relations between the geometric quantities $\calm$, $\kappa\cala$, $v_H\calp$ and $\calt\call$, where the quantity $V_H$ is defined below.
The first of these formulas holds for the entire class of spacetimes under consideration.  The second Smarr formula holds under the additional assumption of exact translation invariance in the compact direction.  A linear combination of these two formulas gives the relation between mass and tension for the boosted black string.
We derive each of these Smarr formulas in two ways, first using scaling arguments (as in {\it e.g.} reference \cite{Chowdhury:2006qn}) and second using Komar integral relations (as in reference 
\cite{Harmark:2003dg}).  Finally, we present a Gibbs-Duhem formula that relates variations in the tension to variations in the other intensive parameters.  This result generalizes the `tension first law' of reference \cite{Traschen:2001pb}.

Our result for the first law resolves a small puzzle related to the  boosted black string, which formed part of the motivation for this work.    It was found in reference \cite{Hovdebo:2006jy} that the tension of the boosted black string becomes negative for values of the boost parameter in excess of a certain critical value, which depends only on the spacetime dimension.  If the physical interpretation of the tension based on equation (\ref{static-first-law}) were to continue to hold in the stationary case, then  the energy of the system would  decrease with increasing $\call$, which seems counter-intuitive.  This puzzle is resolved by showing that the coefficient of the $d\call$ term in the  first law for black holes is an effective tension $\hat\calt$.  The effective tension $\hat\calt$ is equal to the ADM tension in the static case, but includes a contribution from the momentum in the stationary case.  For the boosted black string $\hat\calt$ is always positive, and is in fact given by the tension of the unboosted black string with the same horizon radius.

\section{Stationary, non-rotating Kaluza-Klein black holes}\label{assumptions}

We  consider stationary $D$-dimensional vacuum black hole spacetimes that are asymptotic  to 
$M^{D-1}\times S^1$, and assume that the black hole horizon is a bifurcate Killing horizon.   In accordance with our focus on linear momentum around the $S^1$, we take the ADM angular momentum to vanish.   We denote the horizon generating Killing field by $l^a$, and assume that at infinity it has the form
\begin{equation}\label{at-infinity}
l^a=T^a +v_H Z^a
\end{equation}
where  $T^a=(\partial/\partial t)^a$ and $Z^a=(\partial/\partial z)^a$, with $z$ being the coordinate around the $S^1$.
The surface gravity $\kappa$ of the black hole horizon is defined, as usual, via the relation on the horizon
\begin{equation}\label{surfacegravity}
\nabla_a(l^bl_b)=-2\kappa l_a.
\end{equation}

The form (\ref{at-infinity}) of the horizon generating Killing field at infinity resembles the decomposition of the horizon generating Killing field for a rotating, asymptotically flat black hole.  In that case, {\it i.e.} upon replacing $v_HZ^a$ by $\Omega_H\phi^a$, with $\phi$ an azimuthal coordinate, one can show that $T^a$ and $\phi^a$ are themselves Killing vectors \cite{Hawking:1971vc,Hawking:1973uf,Hollands:2006rj}.  The quantity $\Omega_H$ can then be interpreted as the angular velocity of the horizon, and further shown to be constant on the horizon \cite{Hawking:1971vc}.  

Returning to the case of Kaluza-Klein black holes, the situation is quite different.
Already in the static case, solutions exist which are non-uniform in the $z$ direction.  In the stationary case then, it will not generally be the case that $T^a$ and $Z^a$ are Killing vectors.   For localized black holes or non-uniform black strings with velocity around the $S^1$, only the linear combination $l^a$ is a Killing vector.

\subsection{Two commuting Killing fields}\label{twokillingfields}

It is nonetheless useful to separately consider the case in which $T^a=(\partial/\partial t)^a$ and 
$Z^a=(\partial/\partial z)^a$ are two commuting Killing fields, and that the relation (\ref{at-infinity}) holds throughout the spacetime.
The boosted black string falls into this class of spacetimes.  
If both $Z^a$ and $T^a$ are Killing fields, then the quantity $v_H$ in equation (\ref{at-infinity}) may be considered to be the velocity of the black hole horizon.  This identification follows in a similar way to that of $\Omega_H$ as the angular velocity in the rotating case (see {\it e.g.} the article by Carter in \cite{Carter:1987hk}).  It follows from equation (\ref{surfacegravity}), together with our assumption that $T^a$ and $Z^a$ are commuting Killing vectors, that in addition to $l^al_a=0$ on the horizon, one also has there the orthogonality relations
\begin{equation}\label{orthogonal}
l^a T_a =0,\qquad l^a Z_a =0.
\end{equation}
Given these, one can then show that the metric components on the horizon satisfy 
the two relations
\begin{equation}\label{velocity}
(T^a Z_a)^2 = (T^a T_a)\; Z^b Z_b ,\qquad  v_H =- {T^a Z_a\over (Z^bZ_b) }
  \end{equation}
The second of these leads to the interpretation of  $v_H$ as the 
velocity of the horizon in the following manner.
For rotating black holes, one considers `zero angular momentum observers' or ZAMO's.  The angular velocity $\Omega_H$ of the horizon is the limit of a ZAMO's  angular velocity as it approaches the horizon radius.  For a boosted black string, we may analogously consider observers with zero linear momentum along the string, which we might be justified in calling ZELMO's.
Let $p^a =m\; dx^a /d\tau$ be the momentum of a particle following a geodesic.  It's  energy
$E=-T^a p_a$ and the $z-$component of its momentum $P=Z^a p_a$ are both constants of motion. The condition $P=0$ of vanishing linear momentum  is then $dz /dt = - g_{tz} /g_{zz} $, and we see
that on the horizon the coordinate velocity of a ZELMO is
equal to $v_H$.

\section{ADM mass, tension and momentum}\label{ADM}
 
We review  the formulas for the ADM mass, tension and momentum.  Let us write the spacetime metric near infinity as $g_{\ab}=\eta_{\ab}+\gamma_{\ab}$, where $\eta_{\ab}$ is the $D$-dimensional  Minkowski metric.  The components of $\gamma_{\ab}$ are assumed to fall-off sufficiently rapidly that the integral expressions for the mass, tension and momentum are well-defined.  In the asymptotic region, write the spacetime coordinates as $x^a=(t,z,x^i)$, where $i=1,\dots,D-2$ and the coordinate $z$ running around the $S^1$ is identified with period $\call$.  
Let $\Sigma$ be a spatial slice and  $\partial\Sigma_\infty$  its boundary at spatial infinity.  The ADM mass and momentum in the $z$ direction are  then given in asymptotically Cartesian coordinates by the integrals
 \begin{eqnarray}\label{admdefn}
\calm  &= & {1\over 16\pi G}\int_{\partial\Sigma_\infty} dz\ ds_i \left( -\partial^i\gamma_j{}^j-
\partial^i \gamma_z{}^z +\partial_j \gamma^{ij}\right) \\
\calp & = & {1\over 16\pi G}\int_{\partial\Sigma_\infty} dz\ ds_i \, \partial^i \gamma_{tz}
\end{eqnarray}
where indices are raised and lowered with the asymptotic metric $\eta_{\ab}$ and the area element $ds_i$ is that of a sphere $S^{D-3}$ at infinity in a slice of constant $t$ and $z$.  

The ADM tension is similarly given by the integral \cite{Traschen:2001pb}\cite{Townsend:2001rg}\cite{Harmark:2004ch}
\begin{equation}\label{tensiondef}
\calt= - {1\over 16\pi G}\int_{\partial\Sigma_\infty/S^1} \ ds_i \left( -\partial^i\gamma_j{}^j-\partial^i \gamma_t{}^t +\partial_j \gamma^{ij}\right).
\end{equation}
Note that in contrast with the ADM mass and momentum,  the definition of the tension does not include an integral in the $z$-direction.     
The ADM mass is an integral over the boundary of a slice of constant $t$, which includes the direction around the $S^1$.   The tension, on the other hand,  is defined \cite{Traschen:2001pb}\cite{Townsend:2001rg}\cite{Harmark:2004ch} by an integral over the boundary of a slice of constant $z$.  This includes, in principle,  an integration over time.    However, if one expands the integrand around spatial infinity, one finds that terms that make non-zero contributions to the integral are always time independent.  Time dependent terms fall-off too rapidly to contribute.  Hence, one can omit the integration over the time direction and work with the quantity $\calt$ defined above, which is strictly speaking a  `tension per unit time'.  

We can evaluate these formulas for $\calm$, $\calp$ and $\calt$ in terms of the asymptotic parameters of our spacetimes.
The  spacetimes we consider have topology $R^{D-1}\times S^1$, the coordinate $z$ in the compact direction being identified with period $\call$.  We can write the metric explicitly as 
\begin{equation}
ds^2 =  g_{tt} dt^2 + 2g_{tz}dt dz +g_{zz}dz^2 +2(g_{ti}dtdx^i +g_{zi}dzdx^i) + g_{ij} dx^i dx^j  
\end{equation}
where $x^i$ with $i=1,\dots D-2$ are the non-compact spatial coordinates.
We assume the following falloff conditions at spatial infinity
\begin{equation}\label{falloff}
 g_{tt} \simeq -1 +c_t / r ^{D-4}  ,\qquad  g_{zz}  \simeq  1 +c_z /r ^{D-4} ,\qquad  
  g_{tz} \simeq  c_{tz} / r ^{D-4},
\end{equation}
and further that the coefficients $g_{ti}$ and $g_{zi}$ falloff sufficiently fast that they do not contribute to any ADM integrals at infinity.
The mass, tension \cite{Harmark:2003eg} and momentum can then be shown, using the field equations, to be given in terms of the asymptotic parameters $c_t$, $c_z$ and $c_{tz}$ by
\begin{equation}\label{massandtension}
{\cal M} = {\Omega_{D-3}\L\over 16\pi G} ( (D-3) c_t -c_z ),\qquad
  {\cal T}  = {\Omega_{D-3}\over 16\pi G} (  c_t -(D-3)c_z ),
\end{equation}
\begin{equation}\label{momentum}
\P = - (D-4){\Omega_{D-3}\L\over 16\pi G}c_{tz}.
\end{equation}

 \section{The boosted black string}\label{boostedstring}
 
 The boosted black string serves as a simple analytic vacuum spacetime in which to check the results we present below for the first law, Smarr and Gibbs-Duhem relations.  The boosted black string metric may be obtained starting from the uniform black string, performing a boost transformation with parameter $\beta$ and identifying the new, boosted $z$ coordinate with period $\call$.   This gives
\begin{eqnarray}
ds^2  = && -(1- {c\over r^{D-4}}\cosh^2\beta)dt^2 + (1+{c\over r^{D-4}}\sinh^2\beta)dz^2  \\
&& +2 {c\over r^{D-4}} \sinh\beta\cosh\beta  dzdt + (1- {c\over r^{D-4}})^{-1}dr^2 +r^2d\Omega^2_{D-3} \nonumber
\end{eqnarray}
The horizon, which has topology $S^{D-3}\times S^1$ is located at  $r_H=c^{1/(D-4)}$.  From the asymptotic form of the metric, one finds using the expressions (\ref{massandtension}) and (\ref{momentum})  that the ADM mass, tension and momentum are given as in \cite{Hovdebo:2006jy} by 
\begin{eqnarray}
\calm &=&  {\Omega_{D-3}\call\over 16\pi G}   r_H^{D-4}  ((D-4)\cosh^2\beta+1)\\
\calt &=& {\Omega_{D-3}\over 16\pi G}   r_H^{D-4} (1-(D-4)\sinh^2\beta)  \\
\calp &=& - {\Omega_{D-3}\call\over 16\pi G}   r_H^{D-4} (D-4)\sinh\beta\cosh\beta
\end{eqnarray}
 Note that, as mentioned in the introduction, the tension becomes negative for $\sinh^2\beta> 1/(D-4)$.  We can further compute, as in reference \cite{Hovdebo:2006jy}, that the horizon area, surface gravity, and horizon velocity of the boosted black string are given by
\begin{equation}
\cala=\Omega_{D-3}\call r_H^{D-3}\cosh\beta,\qquad
\kappa = {D-4\over 2 r_H\cosh\beta},\qquad
v_H = -{\sinh\beta\over \cosh\beta}.
\end{equation}

\section{Gauss' Laws for Perturbations}\label{gausslaw}

Following the work of \cite{Sudarsky:1992ty}, we use the Hamiltonian formalism of general relativity to derive the first law for stationary, non-rotating Kaluza-Klein black holes.  Another of our goals is to derive a `first law' for  variations in the tension as in reference \cite{Traschen:2001pb}, for this class of spacetimes.  This requires a slight generalization of  the Hamiltonian formalism to accomodate evolution of data on timelike surfaces in a spacelike direction.  Although, as we discuss below in section 
(\ref{G-D}), we have not yet succeeded in providing a Hamiltonian derivation of the `tension first law' in the stationary case, our presentation of the Hamiltonian formalism will be general enough to provide the necessary tools.

The essence of the method is as follows.  In vacuum gravity, suppose one has a black hole solution with a Killing field.  Now consider solutions that are perturbatively close to this background solution, but are not required to have the original Killing symmetry.  The linearized Einstein constraint equations on a hypersurface can be expressed in the form of a Gauss' law (see \cite{Traschen:1984bp}), relating a boundary integral at infinity to a boundary integral at the horizon.
The physical meaning of this Gauss' law relation depends on the choice of Killing field, as well as on the choice of hypersurface.   Taking the generator $l^a$ of a Killing horizon, together with an appropriate choice of a spacelike hypersurface, yields the usual first law for variation of the mass \cite{Sudarsky:1992ty}.  In the case of solutions that are $z$ translation invariant, choosing 
the spatial translation Killing vector $Z^a$, again with an  appropriate choice of a timelike hypersurface, gives a `first law' for variations in the tension \cite{Traschen:2001pb}.

The formalism then proceeds in the following way.   Assume we have a foliation of a spacetime by a family of hypersurfaces of constant coordinate $w$.  We denote these hypersurfaces, both collectively and individually, by $V$ and the unit normal to the hypersurfaces by $w^a$.  With the application to tension in mind, we consider both timelike and spacelike normals by setting $w_aw^a=\epsilon$ with $\epsilon=\pm 1$.   This slight generalization introduces factors of $\epsilon$ into a number of otherwise standard formulas.
The spacetime metric can be written as
\begin{equation}\label{metricsplit}
g_{ab} = \epsilon w_a w_b + s_{ab}
\end{equation}
where $s_{ab}$, satisfying $s_a{}^bw_b =0$, is the metric on the hypersurfaces $V$. 
As usual, the dynamical variables in the Hamiltonian formalism  are
the  metric $s_{ab}$ and its canonically conjugate momentum $\pi ^{ab} =
\epsilon\sqrt{|s|} (Ks ^{ab}- K^{ab} )$.  Here $K_{ab} =s_a{}^c \nabla _c w_b$
is the extrinsic curvature of a hypersurface $V$.  We consider Hamiltonian evolution
along the vector field $W^a=(\partial/\partial w){}^a$, which can be decomposed into its components normal and tangential to $V$, according to
$W^a  = Nw^a + N^a$ where $N^a w_a =0$. 
As usual, we refer to $N$ and $N^a$ respectively as the lapse function and the shift vector.
The gravitational Hamiltonian density 
which evolves
the system along $W^a$ is then given by ${\cal H} =NH+N^a H_a$ with
\begin{equation}\label{hamconstraint}
H =  -  R^{(D-1)}  + {\epsilon \over |s|}  ({\pi ^2 \over D-2 } - \pi^{ab} \pi_{ab} )
\end{equation}
\begin{equation}
H_b =-2  D_a (|s|^{-{1 \over 2}} \pi^{ab} ).
\label{momconstraint}
\end{equation}
where $R^{(D-1)}$ is the scalar curvature for the metric $s_{ab}$ and the derivative operator $D_a$ on the hypersurface $V$.
One further finds that the quantities $H$ and $H_a$ are simply related to the normal components of the Einstein tensor,
\begin{equation}
H =2\epsilon\; G_{ab} w^a w^b,\qquad H_b =2\epsilon\; G_{ac} w^a s^c {}_b
\end{equation}
These components of the field equations contain only first derivatives with respect to the coordinate $w$,  and hence represent constraints on the dynamical fields, $s_{ab}$ and $\pi^{ab}$, on $V$.  
This property is independent of whether the normal direction is timelike, as in the usual ADM formalism, or spacelike.
In vacuum, the equations $H=0$ and $H_b=0$ are enforced in the Hamiltonian formalism as the equations of motion of the nondynamical lapse and shift variables. These  are referred to as the Hamiltonian and momentum constraints,  a terminology we continue to use in the case that the normal $w_a$ is spacelike.



Let us now assume that the spacetime metric $\gzero$ is a  solution to the vacuum Einstein equations\footnote{In this paper we will focus on the case when the background spacetime is vacuum. It is straightforward to add stress-energy which is described by a Hamiltonian \cite{Traschen:2001pb}. If the matter Hamiltonian contains constraints--such as for Maxwell theory--then 
additional charges appear in the
first law. This was worked out for Einstein-Yang-Mills in \cite{Sudarsky:1992ty}. 
The general case when the background spacetime has  stress-energy, such as a cosmology, was studied earlier in \cite{Traschen:1984bp}.  In this situation, the criterion for a Gauss' law on perturbations is that the background have an Integral Constraint Vector.}
with a  Killing vector
$\xi^a$.
We  decompose $\xi^a$ into components normal and tangent to the hypersurfaces $V$ introduced above, according to $\xi^a=Fw^a +\beta^a$.
Now, let us further assume that the metric $g_{ab} =\gzero +\delta g_{ab}$ is the linear approximation to 
another solution to the vacuum Einstein equations.   
Denote the Hamiltonian data for the background metric by $\szero ,\pizero$, the corresponding perturbations to the data by $h_{ab}=\delta s_{ab}$ and $p^{ab}=\delta \pi ^{ab} $, and the linearized Hamiltonian and momentum constraints by $\delta H$ and $\delta H_a$.   
As shown in \cite{Traschen:1984bp,Sudarsky:1992ty,Traschen:2001pb}, the following statement then holds as a consequence of Killing's equation in the background metric,
\begin{equation}
F\delta H + \beta^a\delta H_a= -\bar D_a B^a
\end{equation}
where $\bar D_a$ is the background 
covariant derivative operator on the hypersurface and  the vector $B^a$ is given by
\begin{equation}\label{gaussvector}
 B^a =   F(\bar D^a h - \bar D_b h^{ab})  - h \bar D^a F + h^{ab} \bar D_b F 
+ {1 \over \sqrt{|\bar s|} } \beta^b(\bar \pi^{cd} h_{cd} \bar s^a{}_b - 2\bar \pi^{ac} h_{bc} -2p^a{}_b ).
 \end{equation}
Here indices are raised and lowered with the background metric $\bar s_{ab}$.  
 Since the metric $g_{ab}$ solves the vacuum Einstein equations by assumption, we know that 
$\delta H=\delta H_a=0$.   Therefore, we have the Gauss'  law type statement $\bar D_a B^a = 0$.
Note that the detailed form of this relation for the perturbations $h_{ab}$ and $p^{ab}$ depends on the 
the Killing vector $\xi^a$ and the normal $w_a$ to the hypersurface, as well as on the background spacetime metric.   Making different choices for the Killing vector and normal can lead to different relations of this form.
We can now integrate the relation $\bar D_a B^a = 0$ over the hypersurface $V$ and use Stokes theorem 
 to obtain
 \begin{equation}\label{gaussint}
 \int  _{\partial V} da_c B^c = 0,
 \end{equation}
 where for black hole spacetimes the boundary $\partial V$ of the hypersurface $V$ typically has two components, one on the horizon and one at infinity\footnote{Kaluza-Klein bubble spacetimes, which we do not consider here,  provide an interesting contrast . There is no interior horizon, but the rotational Killing field has an axis. Hence to use Stokes theorem, one must exclude the axis, which introduces an inner boundary.}.
 
 \section{The first law for stationary, Kaluza-Klein black holes}
 
Following references \cite{Sudarsky:1992ty,Kastor:2006ti}, we now use the Hamiltonian formalism presented in the last section to derive the first law for stationary, non-rotating Kaluza-Klein black holes.  The first  law relates the difference $\delta\A$ in the horizon area  between nearby solutions to the variations $\delta\M$,  $\delta\P$ and $\delta\L$ in the mass, momentum  and length of the compact direction.
As in reference \cite{Kastor:2006ti}, we carry out the calculation first holding the length at infinity, $\L$, fixed, and then use this result in order to do the calculation with $\delta\L\neq 0$.
   
 We assume as in section (\ref{assumptions}) that we have a stationary, non-rotating Kaluza-Klein black hole solution with metric $\gzero$ and horizon generating Killing field $l^a$, which is given  at infinity by  $l^a=T^a+v_H Z^a$. We further assume as in section (\ref{gausslaw}) that the metric 
 $g_{ab}=\gzero +\delta g_{ab}$ is a linear approximation to a solution of the field equations.   At this stage, we assume that $\delta g_{ab}$ is such that $\delta\L=0$.  Further on,  we will relax this assumption.
 
 The derivation of the mass first law is then quite similar to that for rotating black holes \cite{Sudarsky:1992ty}.   Consider a spacelike hypersurface $V$, which intersects the horizon at the bifurcation surface and has a unit normal approaching the vector $T^a$ at infinity.   Choose the Killing vector in the Gauss' law construction to be the horizon generator $l^a$.  
Let $\partial V_\infty$ and $\partial V_H$ denote the boundaries of the hypersurface $V$ at infinity and at the horizon bifurcation surface.  Equation (\ref{gaussint}) then implies that
\begin{equation}\label{basiclaw}
I_H+I_\infty =0
\end{equation}
where
\begin{equation}
I_H=\int_{\partial V_H} da_c B^c,\qquad I_\infty=\int_{\partial V_\infty} da_c B^c.
\end{equation}

Let us first consider the calculation of $I_H$.
On the horizon bifurcation surface, the quantities $F$ and $\beta^a$ vanish, and  the boundary integral on the horizon reduces to
\begin{equation}
 I_H= -  \int_{\partial V_H} da \hat\rho_c (-h\, \bar D^cF + h^{cb}\, \bar D_b F)
\end{equation}
where $ \hat\rho_c$ is the unit outward pointing normal to the bifurcation surface within $V$.  One can show that the surface gravity is given by $\kappa=\hat\rho^c\partial_c F$, and it then follows as in reference \cite{Sudarsky:1992ty} that 
\begin{equation}\label{horizonterm}
I_H = 2\kappa\,\delta\A
\end{equation}

Now consider the boundary term at infinity.  Many of the terms in (\ref{gaussvector}) fall off too rapidly to make non-zero contributions.   In particular, it is straightforward to check that the $D_aF$ terms, as well as those including products of $\bar\pi^{ab}$ with the metric perturbation, fall off too rapidly as $r\rightarrow\infty$ to contribute.  Furthermore, it is sufficient to take $F\simeq 1$ and $\beta^z=v_H$
in this limit.
We then arrive at the expression
\begin{equation}\label{stillneedswork}
I_\infty = \int_{\partial V_\infty} dz\; ds_i (\partial^i h -\partial_j h^{ij}
- 2v_H p^i{}_z)
\end{equation}
At this point, we need to note that the formulas  (\ref{admdefn}) and (\ref{tensiondef}) 
for the ADM mass, momentum and tension are written in terms of the variable $\gamma_{ab}$ defined by $g_{ab}=\eta_{ab}+\gamma_{ab}$.   In order to interpret the boundary integral (\ref{stillneedswork}) in terms of variations in $\calm$, $\calp$ and $\calt$, we need to relate the perturbations $\gamma_{ab}$ and $p^{ab}$ in the Hamiltonian formalism to a covariant perturbation $\delta \gamma_{ab}$.  It is straightforward to show that to the required order of accuracy 
$h=\delta^{kl}\delta\gamma_{kl} +\delta\gamma_{zz}$
and 
$h^{ij}\simeq \delta^{ik}\delta^{jl}\delta\gamma_{kl}$, 
while  a further  calculation reveals that $p^i{}_z\simeq -(1/2)\, \partial^i \delta h_{zt}$.  We then find that 
\begin{eqnarray}
I_\infty &=& \int_{\partial V_\infty}  da_i\left\{ \partial^i h-\partial_j h^{ij} 
+v_H\partial^i h_{zt}\right\} \\
&=&-16\pi G(\delta\M - v_H\delta\P)
\end{eqnarray}
%
Inserting these results into equation (\ref{basiclaw}) then yields the mass first law for boosted black strings (with the length $\L$ at infinity held fixed)\footnote{We can also include perturbative  stress energy in this relation, in which case the mass first law becomes
 \begin{equation}\label{deltamass}
 \delta\mass= {\kappa\over 8\pi G} \delta A +v_H\delta\P +\int_{V} (-\delta T^a _b n_a l^b).
 \end{equation}
 }
\begin{equation}\label{without}
\delta\M={\kappa\over 8\pi G}\delta\A+v_H\delta\P
\end{equation}
We see that the momentum appears as an extensive parameter in the first law, while $v_H$, which for the boosted black string is the horizon velocity, appears as an intensive parameter.  This parallels the way angular momentum enters the first law for rotating black holes.  Equation (\ref{deltamass}) is easily verified for the case of the boosted black string using the formulas of section (\ref{boostedstring}).

We now generalize the first law (\ref{without}) to include perturbations with $\delta\call\neq 0$. Our analysis of the boundary term is based on that in \cite{Kastor:2006ti} for the static case.  
The boundary integral at the horizon in this case remains unchanged and is still given by equation 
(\ref{horizonterm}).  
Additional terms, however, occur  in  the boundary term at infinity. 
Given the results above, we can write the boundary term at infinity as
\begin{equation}\label{begining}
I_\infty = 16\pi G( - \delta\calm  |_{\delta\call=0} +v_H
  \delta\calp |_{\delta\call=0} +\lambda\delta\call ), 
\end{equation}
where $\lambda$ remains to be determined.  On the other hand, we know the $\call$ dependence of $\calm$ and $\calp$ explicitly from the expressions (\ref{massandtension}) and (\ref{momentum}).  This allows us to write
\begin{eqnarray}
\delta\calm &=& \left .\delta\calm\right|_{\delta\call=0} + {\calm\over\call}\delta\call    \\
\delta\calp &=& \left . \delta\calp\right|_{\delta\call=0} + {\calp\over\call}\delta\call 
\end{eqnarray}
Combining these with equations (\ref{begining}), (\ref{basiclaw}) and (\ref{horizonterm}) then gives
\begin{equation}\label{gettingthere}
I_\infty = 16\pi G\left(
-\delta\calm + v_H\delta\calp +(\lambda+{\calm\over\call}-v_H{\calp\over\call})\delta\call \right).
\end{equation}
We can now further appeal to the results of \cite{Kastor:2006ti} for the case $\calp=0$ and write 
$\lambda= \lambda |_{\calp=0} +\lambda^\prime$.  We know from \cite{Kastor:2006ti} that  
$\lambda |_{\calp=0} +\calm/\call=\calt$.  Putting this together, allows us to rewrite (\ref{gettingthere}) as
\begin{equation}
I_\infty = 16\pi G\left(
-\delta\calm + v_H\delta\calp +(\lambda^\prime+\calt -v_H{\calp\over\call})\delta\call \right).
\end{equation}

We still need to calculate the quantity $\hat I_\infty = 16\pi G\lambda^\prime\delta\call$
which includes only the terms in $I_\infty$ that are proportional to both
 $\calp$ and $\delta\call$.  
It is noted in \cite{Kastor:2006ti} that in order for  the perturbative Gauss's law  (\ref{gaussint}) to apply  with $\delta\call\neq 0$, one need to make a coordinate transformation so that $\delta\call$ appears in the metric perturbation, rather than in a change in the range of coordinates.  Following this procedure yields the metric perturbations
\begin{equation}\label{variations}
h_{zz}\simeq 2{\delta\call\over\call}
(1+{c_z\over r^{D-4}}),\qquad
h_{zt}\simeq {\delta\call\over\call}{c_{tz}\over r^{D-4}}
\end{equation}
There are two terms in equation (\ref{gaussvector}) that  potentially contribute to $\hat I_\infty$ and we accordingly write $\hat I_\infty=  \hat I_\infty^{(1)}+\hat I_\infty^{(2)}$.  The first of these terms is given by
 \begin {eqnarray}
 \hat I_\infty^{(1)} &=&   \int _{\partial V_{\infty}}dz da_c{-2\beta^b\bar \pi^{ac}h_{ab} \over \sqrt{|\bar s|}} \\
 &=& \int _{\partial V_{\infty}}dz da_i (-2v_H\bar\pi^{iz}h_{zz})\\
 &=& \int _{\partial V_{\infty}}dz da_i  \, v_H\partial_i\bar g_{tz} \,{2\delta\call\over\call} \\
 &=& 16\pi G\;  v_H\; \calp {2\delta\call\over\call}
 \end{eqnarray}
 The second term, which requires some care in evaluating,  is given by
 \begin{eqnarray}
 \hat I_\infty^{(2)} & = & \int _{\partial V_{\infty}}da_c{-2\beta^b\delta\pi^c_b \over \sqrt{|\bar s|}} \\ 
 &=&  \int _{\partial V_{\infty}}da_i(-2 v_H p^i{}_z)\\ \label{needsexplanation}
   &=&  \int _{\partial V_{\infty}}da_i v_H \; \partial_i\bar g_{tz}\; {\delta\call\over\call}(1-2+1)\\
   &=& 0.
 \end{eqnarray}
 where the factor $(1-2+1)$ in line (\ref{needsexplanation}) comes about in the following way.
 We have $p^{iz}\simeq\delta(\sqrt{s} s^{zz} K_{iz})$ with $K_{iz}\simeq -(1/2)\partial_i g_{tz}$.  The first $1$ comes from the variation of the volume element $\sqrt{s}$, the $-2$ comes from the variation of  inverse metric component $s^{zz}$ following from  equation (\ref{variations}), and the final  $1$ comes from the variation of $g_{tz}$, also as in equation (\ref{variations}).
Putting these results together gives
$\lambda^\prime = 2 v_H\calp/\call$ and hence
\begin{equation}
I_\infty = 16\pi G\left(
-\delta\calm + v_H\delta\calp +(\calt + {v_H\calp\over\call})\delta\call \right).
\end{equation}
Finally, combining this with $I_H$ gives the first law
\begin{equation}\label{firstlaw}
\delta\calm= {\kappa\over 8\pi G} \delta\cala +v_H\delta\calp +(\calt + {v_H\calp\over\call})\delta\call 
\end{equation}
From the $\delta\call$ term, we see that the coefficient of $\delta\call$ is an effective tension given by $\hat\calt=\calt+v_H\calp/\call$.   As mentioned in the introduction, the tension of the boosted black string
becomes negative for sufficiently large boost parameter.  It is straightforward to check that the first law 
(\ref{firstlaw}) is satisfied for variations within the family of boosted black string solutions in section (\ref{boostedstring}), and also that the effective tension is given by 
\begin{equation}
\hat\calt = {\Omega_{D-3}\over 16\pi G}r_H^{D-4}
\end{equation}
which is equal to the tension of the unboosted black string having the same horizon radius.


\section{Smarr formulas, scaling relations and Komar integrals}\label{smarrformulas}

Smarr formulas are relations between the thermodynamic parameters that hold for black hole solutions that have  exact symmetries.  In this section
we will derive the Smarr formula for stationary, but non-rotating Kaluza-Klein black holes.   We will also derive a second Smarr-type formula that holds in the case of exact translation invariance in the $z$-direction, {\it e.g.} for the boosted black string.  We present two approaches to deriving these formulas.  The first is based on general scaling relations, which are familiar from classical thermodynamics, and the second is based on Komar integral relations.

Given the statement of the first law (\ref{firstlaw}) for stationary, non-rotating Kaluza-Klein black holes, the Smarr formula can be derived by making use of a simple scaling argument (see {\it e,g,} \cite{Chowdhury:2006qn}).   Given any stationary vacuum solution to Einstein's equations, we may obtain a one parameter family of solutions by rescaling all the dimensionful parameters in the given solution in an appropriate way.    If a parameter
$\mu$ has dimensions $(length)^n$, we replace it with $\lambda^n\mu$.   The parameters $c_t$, $c_z$ and $c_{tz}$ that specify the asymptotics of the stationary solution all scale as $(length)^{D-4}$.  If we rescale these accordingly, and also replace $\call$ with $\lambda\call$, then the mass and momentum rescale as 
\begin{equation}
\calm=\lambda^{D-3}\bar\calm,\qquad \calp=\lambda^{D-3}\bar\calp\end{equation}
where $\bar\calm$ and $\bar\calp$ are the mass and momentum  of the original solution.  Similarly, the area of the event horizon of the family of spacetimes will be 
$\cala=\lambda^{D-2}\bar\cala$.  Now consider how these quantities change under a small change in $\lambda$.  We have
\begin{equation}
d\calm= (D-3) \calm {d\lambda\over\lambda},\quad d\calp=(D-3) \calp {d\lambda\over\lambda},
\quad d\cala= (D-2) \cala {d\lambda\over\lambda}, \quad d\call=\call {d\lambda\over\lambda}
\end{equation}
The first law (\ref{firstlaw}) must hold in particular for this variation in $\lambda$.  This will implies that
\begin{equation}\label{first-smarr}
(D-3)\calm=(D-2){1\over 8\pi G}\kappa\cala + \hat\calt\call +(D-3)v_H\calp
\end{equation}
which is the Smarr formula for stationary, non-rotating Kaluza-Klein black holes.
Note that via the scaling argument, the effective tension $\hat\calt$ naturally enters the Smarr formula as well as the first law.

We now derive a second Smarr formula that holds only for solutions, such as the boosted black string,  that have exact translation invariance in the $z$-direction\footnote{It appears likely that the sboosted black strings can be shown to be the only stationary, non-rotating, $z$ translational vacuum solutions with non-singular horizons (see reference \cite{Lee:2007bi}).}.  Note that the mass, momentum and horizon area are all extensive quantities in the compactification length $\call$ and that different values of $\call$ give another one parameter family of solutions.  Within this family we have
\begin{equation}
d\calm = \calm {d\call\over\call},\quad d\calp = \calp {d\call\over\call},\quad  d\cala = \cala {d\call\over\call}
\end{equation}
under a small variation in $\call$.
For the first law to be satisfied under such  variations, we must have
\begin{equation}\label{second-smarr}
\calm={1\over 8\pi G}\kappa\cala + \hat\calt\call +v_H\calp
\end{equation}
Because of the simple extensivity of $\calm$, $\calp$, $\cala$ and $\call$ itself in the length $\call$ of the compact direction, this second Smarr formula takes the form of the usual Euler relation for a thermodynamic system, without any additional dimension dependent prefactors.  Note that by taking a linear combination of the two Smarr formulas (\ref{first-smarr}) and (\ref{second-smarr}), the horizon area term may be eliminated, giving
\begin{eqnarray}
\calm &=& (D-3)\hat\calt\call + v_H\calp \\
&=& (D-3)\calt\call +(D-2) v_H\calp
\end{eqnarray}
For $\calp=0$ this is the well known relation between the mass and tension for a uniform black string.

The Smarr formulas may also be derived by geometrical means using Komar integral relations.  This is done in reference \cite{Harmark:2003dg} for the first Smarr formula in the case $\calp=0$.
For a vacuum spacetime with a Killing vector $k^a$, and a hypersurface $\Sigma$ with boundaries $\partial \Sigma_\infty$ at infinity and $\partial \Sigma_H$ at the black hole horizon,  the Komar integral relation implies the equality $I_{\partial \Sigma_\infty}=I_{\partial \Sigma_H}$ where
\begin{equation}
I_S= -{1\over 16\pi G}\int_S dS_{ab} \nabla^ak^b.
\end{equation}
The first Smarr formula results from taking $k^a$ to be the horizon generator $l^a$ and $\Sigma$ to be a spacelike hypersurface with normal $dt$ at infinity.  The computation of the horizon boundary term in this case is by now quite standard (see \cite{Bardeen:1973gs}).   The horizon generator $l^a$ is null on the horizon and consequently normal to the boundary $\partial \Sigma_H$.  Let $q^a$ be the second null vector orthogonal to $\partial \Sigma_H$, normalized so that $l^a q_a=-1$.  One then has on the boundary $dS_{ab}=2l_{[a}q_{b]}dA$, where $dA$ is the surface area element.  It then follows that
\begin{equation}
I_{\partial \Sigma_H}= {1\over 8\pi G} \kappa\cala
\end{equation}
where we have made use of the definition (\ref{surfacegravity}) of the surface gravity.
The boundary term at infinity may be straightforwardly computed using the asymptotic form of the metric in (\ref{falloff}) and the expressions (\ref{massandtension}) and (\ref{momentum}) for the ADM mass, tension and momentum.  One finds 
\begin{eqnarray}
I_{\partial \Sigma_\infty} &=& {\Omega_{D-3}\call\over 16\pi G} (D-4)c_t \\
&=& {1\over D-2}\left( (D-3)\calm-\calt\call\right)-v_H\calp .
\end{eqnarray} 
Equating the two boundary integrals correctly reproduces the first Smarr formula (\ref{first-smarr}).

The scaling argument that led to the second Smarr formula assumed translation invariance in the $z$-direction, {\it i.e.} that $Z^a$ as well as the horizon generator $l^a=T^a+v_HZ^a$ is a Killing vector.  To give a geometric derivation, we additionally assume, as in section (\ref{twokillingfields}), that  the Killing vectors $T^a$ and $Z^a$ commute.  Let us now take the Killing vector in the Komar construction to be $V^a= v_HT^a + Z^a$, which is orthogonal to the horizon generator $l^a$  both at infinity and on the horizon.  We take the hypersurface $\Sigma$ to be timelike, with normal equal to $dz$ at infinity and proportional to $V_a$ at the horizon.  The normal to the horizon within $\Sigma$ is then proportional to the horizon generator $l_a$, and hence the boundary term at the horizon includes the factor
\begin{equation}
l_a V_b\nabla^aV^b = V_aV_b\nabla^a l^b = 0.
\end{equation}
In the first equality the commutivity of the Killing vectors is used and the second equality follows from Killing's equation.  Hence, the boundary term at the horizon $I_{\partial \Sigma_H}$ vanishes for this choice of Killing field and hypersurface.  The boundary term at infinity is again straightforward to compute using the expressions in (\ref{falloff}),(\ref{massandtension}) and (\ref{momentum}).  We find
\begin{eqnarray}
I_{\partial \Sigma_\infty} &=& -  {\Omega_{D-3}\over 16\pi G} (D-4)(c_z+v_H c_{tz})\\
&=&  - {1\over D-2} (\calm/\call- (D-3)\calt)+v_H\calp/\call .
\end{eqnarray}
Equating this with zero then gives the second Smarr formula (\ref{second-smarr}).

\section{Tension first law and Gibbs-Duhem relation}\label{G-D}

A second kind of variational formula for static Kaluza-Klein black holes was derived in reference \cite{Traschen:2001pb}.  This `tension first law' states that 
\begin{equation}\label{tensionfirstlaw}
\call d\calt = - {1\over 8\pi G}\cala d\kappa
\end{equation}
and holds for perturbations that take a static, translation invariant solution into a nearby solution that is stationary, but not necessarily translation invariant.  It applies, for example, to the perturbation between the marginally stable uniform black string and the static non-uniform black string of reference \cite{Wiseman:2002zc}.  In this section, we discuss the thermodynamic context of this formula and conjecture its extension to include $\calp\neq 0$.

We regard the quantities $\calm$, $\cala$, $\call$ and $\calp$ as extensive parameters, while $\kappa$, $\calt$ and $v_H$ are regarded as intensive parameters.   For thermodynamic systems, the first law  relates variations in the extensive parameters, as does equation (\ref{firstlaw}).  In classical thermodynamics a formula relating the variations of the intensive parameters is known as a Gibbs-Duhem relation.  A Gibbs-Duhem relation can be derived from the first law, together with the variation of an Euler formula, such as equation (\ref{second-smarr}).  In the present case, variation of the Euler formula gives 
\begin{equation}
d\calm={1\over 8\pi G}(\kappa d\cala + \cala d\kappa) +\hat\calt d\call +\call d\hat\calt 
+v_H d\calp +\calp dv_H
\end{equation}
 Combining this with the first law then gives the Gibbs-Duhem relation
\begin{equation}\label{gibbs-duhem}
0 = {1\over 8\pi G}\cala d\kappa +\call d\hat\calt +\calp dv_H
\end{equation}
 which reduces to (\ref{tensionfirstlaw}) for $\calp=0$.  
 
 Note, however, that the Euler formula (\ref{second-smarr}) holds only for $z$-translationally invariant solutions, and hence the result above holds only for perturbations that respect this symmetry, {\it i.e.} within the boosted black string family of solutions.   Equation (\ref{tensionfirstlaw}) was derived in \cite{Traschen:2001pb} via the Hamiltonian perturbation methods of section (\ref{gausslaw}), and does not require that the perturbations are invariant under $z$ translations.  We would like to extend the derivation of \cite{Traschen:2001pb} to the stationary non-rotating case, but have not yet accomplished this.
 
 \section{Conclusions}
 
 We have derived various thermodynamic relations for stationary, non-rotating Kaluza-Klein black holes.  As in reference \cite{Kastor:2006ti}, the derivation of the first law required a careful application of Hamiltonian perturbation theory techniques.  
 Perhaps the most interesting aspect of the first law (\ref{firstlaw}) is the appearance of the effective tension 
 $\hat\calt$ which generally differs from the ADM tension.  For the boosted black string, the ADM tension becomes negative for large boost parameter, while the effective tension remains positive.  We note that the gravitational contribution to the ADM tension was shown to be positive for static spacetimes in reference \cite{Traschen:2003jm} using spinorial techniques.  It should be interesting to see what these techniques yield in the stationary case, {\it e.g.} do they prove positivity of the effective tension.
 
Our results concerning the Smarr formulas in section (\ref{smarrformulas}) are also of interest.  In particular, the parallels  between the scaling argument and Komar integral relation derivations are intriguing and can most likely be understood in a more general setting.   Finally, we would like to be able to give a Hamiltonian derivation of the Gibbs-Duhem, or `tension first law', result in equation 
(\ref{gibbs-duhem}).

 \subsection*{Acknowledgments}
The authors would like to thank Roberto Emparan and Henriette Elvang for helpful discussions.
This work was supported in part by NSF grant PHY-0555304.

\end{document}